# FRAMEWORK FOR AN INTEGRATED LEARNING BLOCK WITH CDIO-LED ENGINEERING EDUCATION

**Mouhamed Abdulla, Amjed Majeed**

School of Mechanical and Electrical Engineering Technology,
Sheridan Institute of Technology and Advanced Learning, Toronto, Ontario, Canada

**Meagan Troop**

Educational Development Consultant and Manager, Centre for Teaching and Learning,
Sheridan Institute of Technology and Advanced Learning, Toronto, Ontario, Canada

**ABSTRACT**

As a CDIO collaborating member, Sheridan's School of Mechanical and Electrical Engineering Technology (MEET) maintains a curriculum that is deeply rooted in skills-based learning, experiential learning, and engineering design. In an effort to ensure our graduates are consistently agile and ready for the workforce, we are taking proactive measures to further improve their learning experiences. An important challenge still impeding our students' knowledge acquisition is the perception that program courses have disjointed learning outcomes. In reality, the course map of programs is carefully designed in such a way that technical skills acquired in particular courses gradually build on each other. Despite the traditional existence of prerequisites and co-requisites, the inaccurate view that courses function independently persists among students and, occasionally, among faculty members. One feasible approach to tackle this pedagogical challenge is to combine various courses into an integrated learning block (ILB) having a unified mission and objective. In general, an ILB is formed by the interconnectivity of at least two courses. At Sheridan's School of MEET, we are applying an ILB with three engineering courses offered within the same semester for all of our Bachelor's of Engineering degree programs. The ILB deliverables are based on the design of a chosen engineering system or subunit in a project-based learning (PBL) environment. The rationale of this paper is to share Sheridan's framework for implementing an ILB in engineering programs and to examine the opportunities and challenges related to this type of curriculum design. In particular, we will discuss the methodology by which courses are selected to form an ILB while taking into account their appropriateness for an industry-driven PBL. This will be followed up with some of the strategies that are proposed to evaluate the performance of students in an ILB through formative and summative assessments based on CDIO competencies.

## INTRODUCTION

In Canada, a Bachelor's of Engineering (B.Eng.) degree is obtained once students successfully complete a specialization program of choice over a duration that is typically four years of full-time education. In the Greater Toronto Area (GTA), there are four comprehensive research-based universities that offer this degree: the University of Toronto, York University, Ryerson University and Ontario Tech University. As of recently, Sheridan is the latest school to offer a B.Eng. degree within this region. As a first step, the Mechanical Engineering (PEQAB, 2014) and the Electrical Engineering (PEQAB, 2017) degree programs were proposed to the Postsecondary Education Quality Assessment Board of Ontario for consideration in 2014 and 2017, respectively. Today, both of these engineering discipline have obtained Ministerial consent to offer the B.Eng. degree programs at Sheridan.

What is unique about both of these degree programs is that they were developed with the Conceive, Design, Implement and Operate (CDIO) Initiative (Crawley, Malmqvist, Ostlund, Brodeur, & Edström, 2014) in mind. In fact, Sheridan is the only engineering school within the GTA where CDIO is explicitly embedded in its degree programs (Abdulla, Motamedi, & Majeed, 2019). Practically speaking, this means that our students will focus on analytical skill sets while also devoting nearly half of their education time working on hands-on projects in state-of-the-art labs that are equipped with industry-standard advanced technologies. Setting Sheridan apart from other local universities is the fact that, as a polytechnic, the school focuses on active skills-based learning, experiential learning, project-based learning (PBL), problem-solving techniques, and applied and experiential research. In addition, both Mechanical and Electrical Engineering students are expected to complete a mandatory four month internship with industrial partners following their second year of study, with the option to complete an additional co-op experience following their third year.

Overall, the B.Eng. degree programs at Sheridan consist of fundamental, discipline-related and elective courses in Mechanical and Electrical Engineering. Irrespective of the chosen discipline, enrolled students will have the option to specialize in either power and energy or in mechatronics. The overall duration of the program spans four years that are split into eight semesters. Throughout the program, students are expected to successfully complete 48 courses or the equivalence of 176 credits. It is interesting to highlight that among the Mechanical and Electrical Engineering disciplines, nearly half of the courses in the program map overlap. To be precise, there are 27 common courses across the two disciplines related to fundamental engineering topics, electives and capstone projects. Moreover, within a specific discipline, roughly 82% of courses are identical between the two possible specialization streams. A summary of the B.Eng. programs offered at Sheridan is outlined in Table 1.

Table 1. Macroscopic view of B.Eng. degree programs offered at Sheridan

| Degree Program | | Bachelor's of Engineering (B.Eng.) | |
|---|---|---|---|
| Discipline | | Mechanical Eng. *or* Electrical Eng. | |
| Specialization Stream *within a discipline* | | Power and Energy *or* Mechatronics | |
| Overall Duration | | 4 years (8 semesters) | |
| Overall Courses | *across disciplines* | common courses 27 courses (92 Cr.) | discipline specific courses 21 courses (84 Cr.) |
| | *within a discipline* | common courses 40 courses (144 Cr.) | specialization specific courses 8 courses (32 Cr.) |
| | *total* | 48 courses (176 Cr.) | |
| First Cohort | | Fall 2019 (Mechanical Eng.) *and* Fall 2020 (Electrical Eng.) | |

Excited by the unique nature of these programs, we are taking proactive measures to deliver an exceptional educational program to our learners. Extrapolating from related diploma programs, we foresee that B.Eng. students will fall into the trap of looking at the learning outcomes of each course offered in the program map in a standalone fashion. This compartmentalized perception of the curriculum is utterly problematic. In reality, the program maps are designed in such a way that skill sets, knowledge, and mindsets acquired in courses gradually build on each other. Building on the work of Pace (2017), who introduced the notion of learning bottlenecks – that is, areas where students tend to get "stuck" and "those places in courses where the stream of learning is particularly apt to be obstructed" (p. 19) – we systematically examined places that could potentially emerge as bottlenecks within the curriculum and identified a disjointed perception of the program as a potential challenge. This bottleneck could in fact be addressed by deliberately integrating the curriculum of courses so as to explicitly pull course subjects that may appear disjointed into a cohesive learning block. Evidently, following many data-driven discussions from environmental scans of other programs and with a close examination of this potential bottleneck, we in fact identify the following interrelated curriculum development questions that are worthy of careful investigation:

- Beyond registration restrictions stipulated by prerequisites and co-requisites, do students actually understand the interrelatedness of courses from a technical viewpoint? Namely, what is the link among engineering courses and to what extent does the learning outcomes of one specialized course practically impact another?

- Do learners have an appreciation for and a grasp of the purpose of non-engineering courses (e.g. mathematics, probability and statistics, economics, technical writing) required in an engineering program map? In other words, do they see how these seemingly tangential subjects will inevitably support specialized discipline courses?

- Can students actually connect the relevance of the content acquired in courses to real-world engineering problems, hands-on scenarios and application use-cases? That is, do students actually know the reason and the practical benefit for taking a certain course and how it all fits in the larger scheme of training the engineers of the future?

One way to tackle these questions is by proposing the implementation of an integrated learning block (ILB) in the curriculum. This paper chronicles some of our experience in this initiative by describing Sheridan's framework for implementing an ILB in engineering programs based on CDIO competencies as outlined in the standards in v2.0 (The CDIO Initiative, 2010) and the revisions proposed for v3.0 (Malmqvist, et al., 2019).

**PEDAGOGICAL ACTIONS AND RATIONALE**

A feasible way to tackle the conundrum of disjointed courses is to integrate various subjects into an ILB (Edström, Gunnarsson, & Gustafsson, 2014). ILBs have the advantage of having a unified mission and learning outcomes. In fact, early work on ILBs can be traced back to nearly three decades ago, where integrating engineering curricula was considered to support students in connecting mathematics, science and engineering together (Froyd & Ohland, 2005). Although each course in an ILB is unique in scope, students engaged with the various courses will gain complementary competencies across areas of study. In general, an ILB is formed by the interconnectivity of at least two courses. At Sheridan's School of MEET, we are applying ILBs with three engineering courses offered within the same semester (Rayegani & Ghalati, 2015). Since some courses can be integrated in an easier manner than others, we specifically designed and developed courses that have synergies and interdependencies amongst them. Ultimately, the decisions involved in the selection of engineering courses to integrate and curriculum development were mutually agreed upon following careful intradepartmental discussions. In such a setting, an ILB committee is formed with multiple stakeholders, including: (a) the course leads of target courses, (b) the ILB coordinator, (c) the Associate Dean, and (d) invitees from industry.

Although in Figure 1, we show the course map for the B.Eng. degree program only in Electrical Engineering, a similar setup is also available for Mechanical Engineering. As is obvious, the program spreads over four years, during which time major group projects are undertaken. Indeed, these projects are conducted using CDIO guidelines, of which Sheridan is a member (Zabudsky, Rayegani, & Ghafari, 2014). However, the CDIO competencies are gradually acquired based on instructional scaffolding. To be precise, first year courses will primarily focus on I-O; second and third year courses on D-I-O; and fourth year courses on C-D-I-O guidelines as a whole (Abdulla, Motamedi, & Majeed, 2019). Furthermore, we aim to form at least one ILB of 16 credits in each of the first three years, where the complexity of engineer design evolves from one CDIO-ILB project to another. In order to avoid overwhelming students with major hands-on group work, we decided to include a gap year between each CDIO-ILB project. As a result, these projects are respectively set in term 2 (highlighted in blue), term 4 (highlighted in orange), and in term 6 (highlighted in green). Granted, in the last year (i.e., terms 7 and 8), nine credits are allocated for the capstone project.

|  | Discipline Courses | | | | | | Breadth Courses |
|---|---|---|---|---|---|---|---|
| Term 1 Fall Year 1 | Calculus 1 | Linear Algebra | Fundamentals of Physics 1 | Exploring Engineering | Engineering in Society – Health & Safety | | Composition & Rhetoric |
| Credits: | Credits: 4.0 (3-0-2) | Credits: 4.0 (3-0-2) | Credits: 4.0 (3-2-0) | Credits: 4.0 (2-2-2) | Credits: 3.0 (3-0-0) | 1st CDIO-ILB Project | Credits: 3.0 (3-0-0) |
| Term 2 Winter Year 1 | Calculus 2 P: Calculus 1 | Intro. to Chemistry for Engineers | **Fundamentals of Physics 2** P: Calculus 1 and Fund. of Physics 1 | **Engineering Design and Problem Solving** P: Exploring Engineering | **Computer Programming** | | Breadth Elective Course |
| Credits: | Credits: 4.0 (3-0-2) | Credits: 4.0 (3-2-0) | Credits: **4.0** (3-2-0) | Credits: **4.0** (3-2-0) | Credits: **4.0** (3-2-0) | | Credits: 3.0 (3-0-0) |
| Term 3 Fall Year 2 | Differential Equations P: Calculus 2 | Fundamentals of Digital Systems | Algorithms and Data Structures P: Computer Programming | Electronic Circuits 1 P: Fund. of Physics 2 | Electrical Circuits and Power P: Fund. of Physics 2 | 2nd CDIO-ILB Project | Breadth Elective Course |
| Credits: | Credits: 4.0 (3-0-2) | Credits: 4.0 (3-2-0) | Credits: 4.0 (2-2-2) | Credits: 4.0 (3-2-0) | Credits: 4.0 (3-2-0) | | Credits: 3.0 (3-0-0) |
| Term 4 Winter Year 2 | Numerical Methods P: Differential Equations | **Signals and Systems 1** P: Diff. Equations | Fundamentals of EM Fields P: Fund. of Physics 2 | **Electronic Circuits 2** P: Electronic Circuits 1 | **Microprocessor Systems** P: Fund. of Digital Systems | Co-op and Career Preparation | Breadth Elective Course  1 year |
| Credits: | Credits: 4.0 (3-0-2) | Credits: **4.0** (2-2-2) | Credits: 4.0 (3-0-2) | Credits: **4.0** (3-2-0) | Credits: **4.0** (3-2-0) | Credits: 1.0 (1-0-0) | Credits: 3.0 (3-0-0) |
|  | Mandatory Internship Term (4 summer months) | | | | | | |
| Term 5 Fall Year 3 | Statistics and Quality P: Linear Algebra | Signals and Systems 2 P: Signals and Systems 1 | Communication Systems P: Signals and Systems 1 | Electric Power Gen. and Transmission P: Electrical Circuits and Power | Introduction to Energy Systems P: Fund. of Physics 2 | 1 year gap | Breadth Elective Course |
| Credits: | Credits: 4.0 (3-0-2) | Credits: 4.0 (3-2-0) | Credits: 4.0 (3-2-0) | Credits: 4.0 (3-2-0) | Credits: 4.0 (3-2-0) | | Credits: 3.0 (3-0-0) |
| Term 6 Winter Year 3 | **Design of Digital Systems** P: Microprocessor Systems | **Control Systems** P: Signals and Systems 2 | **Electric Machines** P: Electrical Circuits and Power | Power Electronics P: Electrical Circuits and Power, and Electronic Circuits 1 | Power Distribution System Design P: Electrical Circuits and Power | Engineering Internship Preparation | Breadth Elective Course |
| Credits: | Credits: **4.0** (3-2-0) | Credits: **4.0** (3-2-0) | Credits: **4.0** (3-2-0) | Credits: 4.0 (3-2-0) | Credits: 4.0 (3-2-0) | Credits: 0.0 (1-0-0) | Credits: 3.0 (3-0-0) |
|  | Optional Co-Op Term (8 to 16 months, year 4) | | | | | | |
| Term 7 Fall Year 4 | Capstone Project (C&D) | Micro-Controller Applications P: Design of Digital Systems | 3rd CDIO-ILB Project | Power System Analysis P: Electric Power Gen. and Trans. | Alternative Energy Systems P: Introduction To Energy Systems | Engineering Economics and Entrepreneurship | Breadth Elective Course |
| Credits: | Credits: 4.0 (2-4-0) | Credits: 4.0 (3-2-0) | | Credits: 4.0 (3-2-0) | Credits: 4.0 (3-2-0) | Credits: 3.0 (3-0-0) | Credits: 3.0 (3-0-0) |
| Term 8 Winter Year 4 | Capstone Project (D, I & O) P: Capstone Project (C&D) | | | Power System Control & Protection P: Power System Analysis | Intelligent Power Systems P: Power System Analysis | | Breadth Elective Course |
| Credits: | Credits: 5.0 (2-6-0) | | | Credits: 4.0 (3-2-0) | Credits: 4.0 (3-2-0) | | Credits: 3.0 (3-0-0) |

Figure 1. Program map for B.Eng. degree in Electrical Engineering for the Power and Energy specialization stream.

**FORMATIVE AND SUMMATIVE FEEDBACK STRATEGY**

In formative assessments, students' performance in a particular skill set emphasized in a specialized course is evaluated by the specific instructor of the course. This is, of course, conducted on a regular basis using different techniques, including traditional written examinations, data collection through hands-on laboratory work, and group reports. On the other hand, the summative assessment of a learning block is evaluated by the ILB committee. This committee will take into consideration the amalgamation of the various acquired skills from the three target courses. They will also identify the means by which these newly acquired skills have helped in the overall engineering design project. In this component, the ILB coordinator will, in fact, have responsibilities quite similar to that of a project manager for an engineering design project. The coordinator will regularly meet with students in order to verify that periodic milestones are successfully met.

The approach by which we assess the performance of students in an ILB formation is structured in a systematic manner (see Figure 2). To be precise, the deliverables expected from an ILB will generally be based on the design of a particular engineering sub-system or system in a PBL environment (Kolmos, 2017). First year projects, such as wind turbine, robotic gripper and spidercam (Germain, 2017), focus on I-O competencies. On the other hand, real-world system engineering design problems geared for D-I-O and C-D-I-O competencies are proposed through close consultation with industrial partners in advanced semesters.

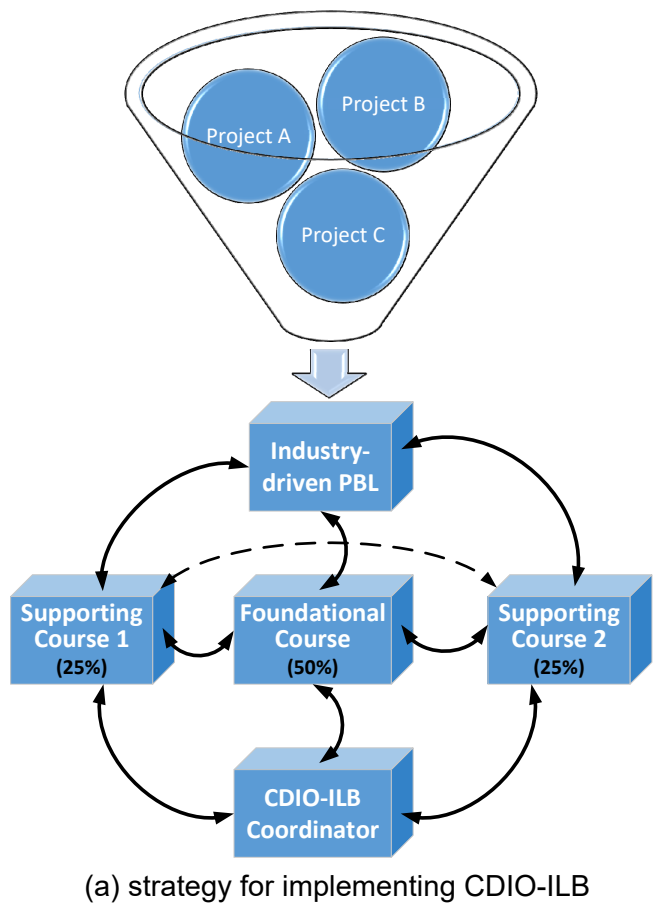
(a) strategy for implementing CDIO-ILB

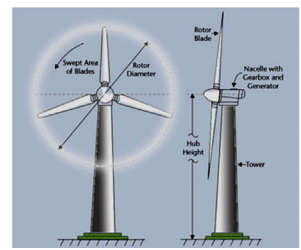
(b) wind turbine project

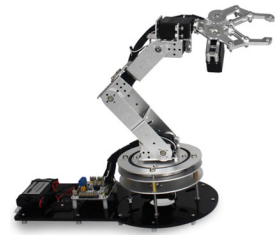
(c) robotic gripper project

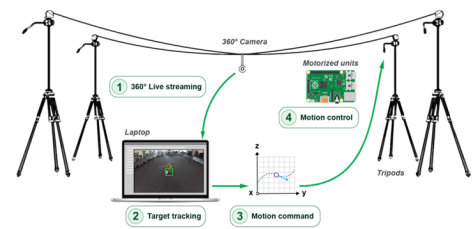
(d) spidercam project

Figure 2. CDIO-ILB setup with three courses founded on PBL, and with sample projects.

Certainly, to ensure the success of an ILB, group formation and interaction is very important. To clarify, we should stress that the same group of students will be registered in similar sections of courses taking part in an ILB. The learners will be responsible to form their groups composed of either 3 or 4 students each. The groups will identify themselves with a name of their choice, and they will remain together across the ILB courses until the end of the semester. The groups will also elect a designated team leader in order to facilitate interaction with instructors.

Evidently, training students in group work is extremely important since effective group interaction is an integral element of a successful professional work environment, and, in general, group work enhances the overall deliverables assigned by a supervisor. Nevertheless, groups are always prone to some challenges. For instance, some group members may not actively participate in the engineering design project. Further, at times, group cohesion and coordination may be lacking and, as a result, efforts will be fragmented. Should this happen, groups will have an intervention with the ILB coordinator in order to suggest techniques to overcome challenges and resolve differences. One mechanism to minimize and safeguard against such possibilities would be to require that each group submit a team contract provided to them at the start of the semester. This is an effective and proven strategy for group harmonization applied in other engineering courses at Sheridan were team work is front and center (e.g., COMM-16165, Technical Reports and Presentations course).

**EARLY RESULTS IN IMPLEMENTING THE CDIO-ILB FRAMEWORK**

As indicated in Table 1, the first cohort of B.Eng. students enrolled in Mechanical Engineering in the fall of 2019. The first CDIO-ILB project began soon after in winter of 2020 during the students' second semester. Meanwhile, in anticipation of the very first ILB experience at Sheridan, the faculty in the School of MEET had regular weekly meetings throughout the fall 2019 semester in order to put concepts of this important framework into action. These meetings were instrumental for promoting extensive exchanges, debates and brainstorming sessions where diverse viewpoints enriched the discussions related to organization and logistics, forecasting potential challenges, and managing and operating PBL activities arranged in an ILB setup. Moreover, in these meetings, we were able to effectively study, refine, and finalize a number of related aspects pertaining to: (a) curriculum mapping to CDIO syllabus and to Canadian Engineering Accreditation Board (CEAB) requirements; (b) evaluation plan and deliverables; (c) rubric design and performance assessments; (e) potential schemes for group size as a function of project complexity; and (f) defining the scope and description of the first set of projects compatible with the learning outcomes of the three courses within an ILB. Of course, estimating budget requirements and mobilizing faculty, engineering staff, and technologists to implement these CDIO-ILB projects was needed to assist our learners engaged in this unique, hands-on undertaking.

As shown in Figure 1, we formed a learning block by integrating the following courses: Engineering Design and Problem Solving (ENGR-18922D), Computer Programming (ENGR-11833D), and Fundamentals of Applied Physics (PHYS-15924D). As illustrated in Figure 2.a, the design course was the foundational course for the CDIO-ILB project, and the programming and physics courses were supporting courses in this PBL assessment. Furthermore, twelve design teams were formed composed of three students each. The groups had the choice to work on one of three projects, shown in Figures 2.b (wind turbine), 2.c (robotic gripper), and 2.d (spidercam). In order to avoid a higher frequency of students working on a particular project over others, we decided to equally divide them based on a first-come, first-served basis.

Groups had the choice to claim their preferences by submitting a form to the CDIO-ILB coordinator indicating their first, second, and third choices. Based on the time log of the submitted form, a project was allocated to a particular group.

Despite the technical nature of this learning activity, engaging in a CDIO-ILB project promoted vital skills needed in the toolkit of future professional engineers, such as project and time management, critical thinking, creativity, and innovation. It also incited students to take proactive measures to seek feedback and suggestions from faculty and subject matter experts in order to improve and polish their project output. Although such attributes are generally seen with senior students engaged in final year capstone projects, it was inspiring and refreshing to see such professional growth among our first year junior students. In fact, the observed response of our students to the CDIO-ILB projects is perfectly aligned with our school's strategy and vision to support our learners in thriving and unleashing their full potential to succeed academically and beyond.

**REFLECTION AND SCHOLARSHIP**

As noted earlier, we are putting in place an evidence-informed framework for implementing an ILB for Sheridan's engineering programs. To the best of our knowledge, very few schools have experimented with an ILB and, even if they have, they generally formed a block based on two courses (Leone & Isaacs, 2001) or have potentially considered an ILB formed with non-engineering courses (Shetty, et al., 2001). At the School of MEET, our vision is to go beyond that and to truly offer a revolutionized curriculum that will in essence prepare our students for the workplace. In other words, as soon as they graduate, we want them to hit the ground running in their respective professional contexts. We want our learners to be competent in technical skill sets, in problem solving techniques, and in having the agility to connect diverse intellectual elements so as to solve a real-world engineering challenge. We also want our students to be professional engineers adept in soft skills which include: technical writing, technical presentations, group harmonization, conflict resolution techniques, and engineering design (Abdulla & Shayan, 2013). If we are successful in this vision, we truly believe that our students will not only be able to find competitive work opportunities, but they will also be prepared to spin off some of their engineering design ideas that incubated at Sheridan.

Since we are in uncharted territory, we anticipate that there will be challenges with the implementation of this curriculum in the next couple of years, both for the learners and for the committee. However, with these challenges, we will have the opportunity to experiment and innovate with various pedagogical alternatives. To this end, in order to extend our research in curriculum design work, we are particularly interested to further investigate the following topics:

- In an eight semester program, how often should a CDIO-ILB project be applied? Is having three ILBs in the program map sufficient or excessive (see Figure 1)? Namely, how do we ensure that we achieve the learning outcomes of knowledge acquisition through an ILB in an adequate and balanced manner?

- How do we ascertain that real-world engineering projects proposed by industry are appropriate for PBL and CDIO (Edström & Kolmos, 2014)? In other words, are the proposed projects compatible with the learning content and outcomes of ILB courses? Will our students have the necessary skills and know-how to embark on these hands-on technical projects?

- Generally a combination of backward (Wiggins & McTighe, 2005) and forward curriculum design (Abdulla, Motamedi, & Majeed, 2019) is applied to a specific course. Can the same methodology be applied to a block of engineering courses in an ILB setup?

**CONCLUSION**

Our prime goal in exploring the CDIO-ILB approach is to ensure that we offer modern and relevant curricula that prepare our students to solve the major complexities of the future. Undoubtedly, the capacity to recognize and connect diverse technical elements to tackle a specific challenge is a vital skill for engineers. This scholarly examination allowed us to look more systematically at the intended student learning experience coupled with CDIO competencies based on instructional scaffolding. Following an elaboration on the means to manage, coordinate and assess the worthiness of CDIO-ILB projects, we highlighted early results in implementing this framework for the first time in this academic year. As we explore, engage and gain experience and feedback from faculty and students involved in ILB, we aim to continue sharing our methodology, framework and data sets with the wider community of engineering educators.

**ACKNOWLEDGMENTS**

This pedagogical research project was supported and funded by the School of Mechanical and Electrical Engineering Technology, Sheridan Institute of Technology, Toronto, Canada. The authors would also like to acknowledge faculty members from the School of MEET and the Centre for Teaching and Learning for providing thoughtful feedback and valuable input as a comprehensive framework for an integrated learning block with CDIO guidelines was drafted.

**REFERENCES**

Abdulla, M., & Shayan, Y. R. (2013). On the Peculiarities of Design: An Engineering Perspective. *Proc. of the 4th Conference on Canadian Engineering Education Association - CEEA'13*, (pp. 1–5). Montréal, QC, Canada. doi:10.24908/PCEEA.V0I0.4823

Abdulla, M., Motamedi, Z., & Majeed, A. (2019). Redesigning Telecommunication Engineering Courses with CDIO geared for Polytechnic Education. *Proc. of the 10th Conference on Canadian Engineering Education Association - CEEA'19*, (pp. 1–5). Ottawa, ON, Canada. doi:10.24908/PCEEA.VI0.13855

Crawley, E. F., Malmqvist, J., Ostlund, S., Brodeur, D. R., & Edström, K. (2014). *Rethinking Engineering Education: The CDIO Approach.* New York, NY: Springer.

Edström, K., & Kolmos, A. (2014, Sep.). PBL and CDIO: Complementary Models for Engineering Education Development. *European Journal of Engineering Education, 39*(5), 539-555.

Edström, K., Gunnarsson, S., & Gustafsson, G. (2014). Integrated Curriculum Design. In *Rethinking Engineering Education: The CDIO Approach* (pp. 85–115, Ch. 4). New York, NY: Springer.

Froyd, J. E., & Ohland, M. W. (2005). Integrated Engineering Curricula. *Journal of Engineering Education, 94*(1), 147–164.

Germain, H. (2017). *Autonomous Spidercam Motion Control* (Master's thesis, University College London, London, UK).

Kolmos, A. (2017). PBL Curriculum Strategies: From Course Based PBL to a Systemic PBL Approach. In *PBL in Engineering Education: International Perspectives on Curriculum Change* (pp. 1–12, Ch. 1). The Netherlands: Sense Publishers.

Leone, D., & Isaacs, B. (2001). Combining Engineering Design With Professional Ethics Using an Integrated Learning Block. *Proc. of the ASEE Annual Conference and Exposition*, (pp. 1–4). Albuquerque, New Mexico, USA.

Malmqvist, J., Knutson Wedel, M., Lundqvist, U., Edström, K., Rosén, A., Fruergaard Astrup, T., . . . Kamp, A. (2019). Towards CDIO Standards 3.0. *Proc. of the 15th International CDIO Conference*, (pp. 44–66). Aarhus, Denmark.

Pace, D. (2017). *The Decoding the Disciplines Paradigm: Seven Steps to Increased Student Learning.* Bloomington: Indiana University Press.

PEQAB. (2014, Dec.). Submission: Bachelor of Engineering (Mechanical Engineering), Applying for Ministerial Consent under the Post-secondary Education Choice and Excellence Act, 2000. Toronto, ON, Canada.

PEQAB. (2017, Feb.). *Submission: Bachelor of Engineering (Electrical Engineering), Applying for Ministerial Consent under the Post-secondary Education Choice and Excellence Act, 2000*. Toronto, ON, Canada.

Rayegani, F., & Ghalati, R. (2015). Integration of Math and Science with Sheridan Engineering Program using CDIO Tools. *Proc. of the 11th International CDIO Conference* (pp. 1–4). Chengdu, Sichuan, P.R. China: Chengdu University of Information Technology.

Shetty, D., Leone, D., Alnajjar, H., Keshawarz, S., Nagurney, L., & Smith, L. T. (2001). Integrating Engineering Design with Humanities, Sciences and Social Sciences Using Integrative Learning Blocks. *Proc. of the ASEE Annual Conference and Exposition*, (pp. 1–9). Albuquerque, New Mexico, USA.

The CDIO Initiative. (2010). *The CDIO Standards v2.0.* CDIO Knowledge Library.

Wiggins, G. P., & McTighe, J. (2005). *Understanding by Design* (Expand second ed.). Alexandria, Virginia: Association for Supervision and Curriculum Development.

Zabudsky, J., Rayegani, F., & Ghafari, S. (2014). Sheridan Journey: Shaping Ideal Engineering Programs based on CDIO Approach. *Proc. of the 10th International CDIO Conference* (pp. 1–9). Barcelona, Spain: Universitat Politècnica de Catalunya.

# BIOGRAPHICAL INFORMATION

*Mouhamed Abdulla*, Ph.D., is currently a Professor of Electrical Engineering at Sheridan. Since 2015, he was a Marie Skłodowska-Curie Individual Fellow supported by the European Commission at Chalmers University of Technology in Gothenburg, Sweden. In 2017, he was a Visiting Fellow with the Dept. of Electronic Engineering of Tsinghua University in Beijing and at Nanyang Technological University in Singapore. Until 2015, he was an NSERC Postdoctoral Fellow with the Dept. of Electrical Engineering of the University of Québec. For nearly 7 years, he was with IBM Canada Ltd. as a Senior Technical Specialist. He received, respectively in 2003, 2006, and 2012, a B.Eng. (with Distinction) in Electrical Engineering, an M.Eng. in Aerospace Engineering, and a Ph.D. in Electrical Engineering, all at Concordia University. His pedagogical research focuses on Engineering Design, Course Development and CDIO.

*Meagan Troop*, Ph.D., is currently a Manager with the Educational Development Team of Sheridan's Centre for Teaching and Learning. At Sheridan, Meagan is involved in supporting faculty and in curriculum development. Prior to that role, Meagan worked as an instructional designer at the University of Waterloo. She holds a Ph.D. degree in Curriculum from Queen's University. Her research interests include creative pedagogies, undergraduate and graduate student development, UX design for learning, and the scholarship of teaching and learning.

*Amjed Majeed*, Ph.D., is currently the Associate Dean of the School of Mechanical and Electrical Engineering at Sheridan. Prior to that role, Amjed was a Faculty Member, Program Chair, and Dean of Academic Operations at the "Higher Colleges of Technology" in the UAE. As Program Chair, Amjed oversaw the development of the Mechanical, Electrical and Mechatronics Engineering degree programs and ABET accreditation. As Dean of Academic Operations, he directed and administered different programs in engineering, business, CIS, education and applied media. From 1999 to 2004, Amjed was an Electrical Engineer Designer at ThyessnKrupp Northern Elevator in Toronto. Amjed holds a B.Sc. in Electrical Engineering, an M.Sc. in Engineering, and a Ph.D. in Computing from Charles Sturt University in Australia.

*Corresponding author*

Prof. Mouhamed Abdulla
School of Mechanical and Electrical Engineering, Sheridan Institute of Technology and Advanced Learning,
7899 McLaughlin Road, Brampton, Ontario, Canada, L6Y 5H9
+1 (905) 845-9430, ext. 5464
mouhamed.abdulla@sheridanc.on.ca